\documentclass[aps,pre,preprint,groupedaddress,showpacs]{revtex4}
\usepackage{epsfig}
\usepackage{amsmath} 
\begin{document}

\title{ Homoclinic crossing in open systems: Chaos in periodically perturbed monopole plus quadrupolelike potentials}
\author{P.S. Letelier and A.E. Motter$^*$}
\affiliation {Departamento de Matem\'atica Aplicada -- IMECC, Universidade
 Estadual de Campinas, 13081-970 Campinas, SP, Brazil\\
[Phys. Rev. E {\bf 60}, 3920 (1999)]}


\begin{abstract}

The Melnikov method is applied to  periodically perturbed open
 systems modeled by an inverse--square--law
 attraction center   plus a quadrupolelike term. A
compactification approach that regularizes periodic
 orbits at infinity is introduced. The (modified)
 Smale-Birkhoff homoclinic theorem is used to study transversal
homoclinic intersections. A larger class of open
 systems with degenerated
(nonhyperbolic) unstable periodic orbits after regularization
 is also briefly considered.
\end{abstract}
\maketitle

\section{Introduction}

Since the pioneering work of Poincar\'e \cite{poincare1} in celestial 
mechanics  in which the mathematical basis of
 deterministic chaos in compact phase space systems was laid down,
the study of homoclinic phenomena in closed systems with
hyperbolic
unstable periodic orbits has allowed the understanding
of a rich variety of
nonlinear effects in  physics, chemistry, and
biology \cite{bai}.
Due to its universality, 
models in which unstable periodic orbits
are subjected to small periodic perturbations has  become one of
the main paradigms of
deterministic chaos \cite{arn}. An analytical tool to study
such models is
the Melnikov method \cite{poincare2,melnikov,arnold} in
connection with
Smale-Birkhoff homoclinic theorem \cite{birkhoff,smale},
and  Kolmogorov-Arnold-Moser (KAM) theory in the Hamiltonian
case \cite{moser}.

The Melnikov function describes the transversal distance between
the stable and
unstable manifolds associated to an unstable periodic orbit. Its
 isolated odd zeros  indicate
transversal intersections between these manifolds, and hence the
onset of chaos
\cite{guckenheimer}. Examples of applications of the Melnikov
method in gravitation
are the motion of particles in perturbed two- and
three-dimensional St\"ackel
potentials \cite{gerh1,gerh2}, the chaotic evolution  of
cosmological models \cite{koiller},
the study of orbits around a black hole perturbed by either
gravitational radiation \cite{vieira} or an external quadrupolar shell
\cite{moeckel}, and  the   bounded  motion of 
particles  in a periodically perturbed attractive center described by
a  monopole plus a quadrupolelike potential were  considered 
in Ref. \cite{letelier}.

The Melnikov method has also been  used in many other
branches of physics. We find
examples of applications of this method to the study of
Josephson junctions
\cite{yao,ken}, planar periodic vortical flows \cite{tae},
solitons
\cite{abd}, liquid crystals \cite{ste}, and transfer dynamics
of quasiparticles
\cite{hen}.

Even for Hamiltonian systems fundamental questions about chaos in
 non-compact phase space systems  remain to be answered. Among the
 more important unsolved  questions are the notion of
chaos itself and the lack of an adequate theory to deal with it. 
 Partial results  obtained in this area are the  fractal techniques
 in scattering processes
\cite{ott}, these are   numerical techniques that
 present some difficulties due to the existence of different 
 time scales  for nearly bounded scattering.  They are inadequate to
the study of   chaotic behavior arising from separatrices between
 bounded and unbounded  orbits. Furthermore, they are unable
to present a complete description of the chaotic motion 
 as the one  provided by  analytical methods in the closed system  case.

The aim of this  paper is to study the homoclinic phenomenon
for a class of  open systems that by a suitable change
of coordinates
can be approached in terms of an adequate formulation
 of the Melnikov method and Smale-Birkhoff homoclinic theorem.
The change of coordinates regularizes
 the unstable periodic orbit at infinity
and it compactifies the region of interest of the phase
 space; however, the phase space
as a whole remains noncompact. Alas the resulting unstable periodic orbit is
 typically nonhyperbolic
and the standard stable manifold theorem, needed to state the
 Melnikov method \cite{guckenheimer}, does not apply. 
  McGehee \cite{gehee} extended this theorem to degenerated
cases in the context of the Newtonian three body problem.
Xia \cite{xia},
 and Dankowicz and Holmes
\cite{dankowicz}, among others, used McGehee's result in
connection with
Melnikov method and Smale-Birkhoff homoclinic theorem
to study the nonintegrability
of the three body problem.

Here we consider the equatorial motion of
a particle moving
 in a potential described
by a monopolar term plus a quadrupolelike contribution. This
 potential models  the gravitational attraction of a
 galaxy bulge or any nonspherical celestial body; it also 
  arises  in general relativity
in the study of the motion of a test particle around a Schwarzschild
 black hole, the quadrupole term being a general relativistic
 effect associated to the angular
 momentum of the particle in the reduced
two-dimensional phase space, see for instance Ref. \cite{chandra}.

In Sec. II  the fixed saddle points associated to the monopole plus
quadrupole system,
as well as the coordinate transformation that regularizes
these points at
 infinity, are studied. In the next section we present some
 mathematical preliminaries
 and the Melnikov method. The equations of motion are used to 
reduce this method to the analysis of simple graphics.
We find that the perturbation induces transverse
 homoclinic orbits in some ranges of
the parameters, and we apply the modified Smale-Birkhoff
 homoclinic theorem to
 verify the presence of a symbolic dynamics equivalent to a Smale horseshoe map;
see Ref. \cite{guckenheimer}.
In Sec. IV the study of the motion is completed with
a presentation of
 Poincar\'e sections  that reveal  different levels of
chaotic behavior as a function of the parameters.
Finally, in the last
 section, we make some remarks
about the class of system in which  the same kind of
analysis can be performed.

\section{The homoclinic orbit and perturbations}

We shall consider the orbit of a particle in a plane
under the influence of a force
modeled by a potential with inverse square law plus
a quadrupolelike term. It is
convenient to work with dimensionless quantities.
The motion of the particle
is described by \cite{letelier}
\begin{equation}
H_0 = \frac{p^2}{2} + \frac{1}{2r^2} -\frac{1}{r}
-\frac{\beta}{r^3}\;\;\; \left( p = \frac{dr}{dt }\right),
\label{1}
\end{equation}
where $r$, $p$, $t$, $H_0$, and $\beta$ are dimensionless
quantities proportional to,
respectively, the radius, the radial momentum, the time,
the Hamiltonian function, and
the quadrupole moment of the attraction center.
The effective potential 
\begin{equation}
V_{eff} = \frac{1}{2r^2} -\frac{1}{r} -\frac{\beta}{r^3}\; ,
\label{2}
\end{equation}
is presented in Fig. 1 for different values of the
parameter $\beta\geq 0$.

The natural
space to study a periodically perturbed planar system
is $I\!\! R^2 \times S^1 $,
where the unstable periodic orbits have a proper meaning \cite{guckenheimer}.
In the corresponding unperturbed autonomous case
the phase space is in $I\!\! R^2$ and the unstable periodic orbits reduce to
fixed saddle points. The unstable periodic orbits are also reduced to
fixed saddle points for the maps defined on Poincar\'e
sections $I\!\! R^2 \times \{ t_0 \} \subset I\!\! R^2 \times S^1$. 

The above system presents a homoclinic loop associated
to the hyperbolic fixed saddle point at $(r,p)=(r_M ,0)$, where
$1/r_M ={1}/{6\beta}+
\sqrt{\left({1}/{6\beta}\right) ^2 -{1}/{3\beta}}$,
for $\beta$ limited by $1/16 <\beta <1/12$.
This case is important
 in the study of bounded orbits
and was explored in Ref. \cite{letelier}.

 In the present work we
 study the instabilities of
 unbounded orbits, the relevant
 values of the parameter are $\beta = 1/16$ and $0\leq\beta <1/16$.
Let $H_0 =0$, if $\beta =1/16$ the points
 $r = r_M = 1/4$ and $r = \infty$ on the $r$ axis
represent fixed points such that the particle takes
 infinity time to reach to or to depart from each of
these points.
For zero energy still, if $0\leq\beta <1/16$ the
motion of the particle is restricted to the
region between $r= r_-$ and $r = \infty$, where $1/r_- = 1/4\beta
 - \sqrt{(1/4\beta)^2 -1/\beta}$,
and only the last point represents a fixed point in this range
of values of $\beta$.

Since the orbits of interest are in a semi-infinity region bounded
 away from the origin,
we can compactify this part of the phase space with a change of the 
 position coordinate like $r=u^\alpha$ with $\alpha<0$.
We find that the
 transformation  $r = 1/u^2$  allows 
 us to model the problem in a way similar to
Refs. \cite{gehee,xia,dankowicz}.
This new coordinate regularizes the fixed point at  infinity 
that now is at the point
$(u,p) = (0,0)$. The zero energy orbits generate a heteroclinic
 loop for $\beta =1/16$
associated to the hyperbolic fixed saddle point at
$(u,p) = (2,0)$ and the
 degenerated fixed saddle point at $(u,p) = (0,0)$,
and a homoclinic loop for $0\leq\beta <1/16$ associated to
 the degenerated fixed saddle point at $(u,p) = (0,0)$.
Degenerated in the sense that both eigenvalue of the
linearized vector
 field are zero, as can
be seen from the Hamiltonian equations
\begin{eqnarray}
&&\frac{du}{dt}=-\frac{1}{2} u^3 p,
\label{3}\\
&&\frac{dp}{dt}= -u^4 +u^6 -3\beta u^8 .
\label{4}
\end{eqnarray}
The points $(u,p)=(2,0)$ and $(u,p)=(0,0)$ correspond to,
respectively, the hyperbolic fixed saddle point at
$(r,p)=(1/4,0)$ for $\beta =1/16$ and the degenerated
fixed saddle point
at $(r,p)=(\infty,0)$ for $0\leq\beta\leq 1/16$.
The homoclinic and heteroclinic loops are defined by
 the intersection
between the stable and unstable manifolds on the $(u,p)$ plane.
The homoclinic loop for $\beta =1/18$ is shown in Fig. 2
 and the heteroclinic one
is presented in Fig. 3.

The explicit  integration of the homoclinic and heteroclinic
loops will be necessary to apply
the Melnikov method and can be obtained from the first
integral of motion ($H_0 =0$).
For $\beta =0$ we find
\begin{equation}
t(v)=\pm \frac{(1+v )(2-v )^{1/2} }{3 v^{2/3}} ,
\label{5}
\end{equation}
where $v= 1/r$, the time origin is take in the symmetry
point of loop, and the sign refers to
the upper ($+$) and lower ($-$) parts of the loop.
 Analogously, for $0<\beta <1/16$ it reads
\begin{eqnarray}
t (v)=
\pm \sqrt{\frac{2}{\beta}}\left\{ \frac{(2v_+ +v_- )F(\delta ,q)
 - 2(v_+ +v_- )E(\delta ,q)}{3v_- ^2 v_+ ^{3/2}}\right. \nonumber\\
\left. +\frac{v_+ v_- + (2v_+ +v_- )v}{3v_+ v_- ^2}\sqrt{\frac{v_-
 -v}{(v_+ -v)v^3}}\right\}
\label{6}
\end{eqnarray}
with
\begin{equation}
\delta = \arcsin\sqrt{\frac{v_+ (v_- -v)}{v_- (v_+ -v)}}, \;\;\;
\mbox{}\;\;\; q = \sqrt{\frac{v_-}{v_+}} \; ,
\label{7}
\end{equation}
where $F(\delta , q)$ and $E(\delta , q)$ are elliptic
integrals of first and second type in the
Legendre normal form (Ref. \cite{gradshteyn}, p. 224), and
$v_\pm = 1/4\beta \pm \sqrt{(1/4\beta) ^2 -1/\beta}$
are the roots of
 $V_{eff}=0$.
Moreover, for $\beta =1/16$ we get
\begin{eqnarray}
t(v ) =
\pm\left\{ \frac{\sqrt{2}}{3}\left( \frac{1}{v^{3/2}} -
\frac{1}{(4/3)^{3/2}}\right) +
\frac{1}{2\sqrt{2}}\left( \frac{1}{\sqrt{v}} -
\frac{1}{\sqrt{4/3}}\right)\right. \nonumber\\
\left. +\frac{1}{8\sqrt{2}}\ln\left[\left( \frac{2-
\sqrt{v}}{2+\sqrt{v}}\right)\left( \frac{2+\sqrt{4/3}}{2-
\sqrt{4/3}}\right)\right]\right\}
\label{8}
\end{eqnarray}
with the choice $t(4/3)=0$, where $v= 4/3$ is the local
minimum of $V_{eff}$.

Now, let us consider the Hamiltonian (\ref{1}) perturbed
by a periodic multipolar term
of the form
\begin{equation}
H = H_0 + \varepsilon H_1 ,
\label{9}
\end{equation}
\begin{equation}
H_1 = r^{-n}\cos (\Omega t)\;\;\; (n\geq 2),
\label{10}
\end{equation}
where $n=2$ is dipolar, $n=3$ is quadrupolar, etc. These
perturbations can model the
attraction  due to a distribution of masses with periodic
motions that are placed
inside the planet  orbit.
  
We shall consider our attraction center with a fix total mass.
In other 
words,
we excluded the monopolar case ($n = 1$) that represents
a periodic variation of the mass.

In the next sections we study  how these perturbations can affect
the dynamics of the system.

\section{Melnikov method}

 Powerful tools to study  near integrable
systems are Melnikov type of techniques that 
 detect transversal intersections
between the stable and unstable manifolds associated to a
unstable periodic orbit. The presence of such transversal intersections
 is a guarantee of
complicated dynamics  and in some cases leads to a symbolic 
dynamics equivalent to the Smale
horseshoe \cite{guckenheimer}.

In order to simplify the analysis it is convenient
 to abstract a little from the particular
problem presented above. In what follows the loops
 are on the $X$ plane and the manifolds
are in the $(X,\theta )$ space, where $\theta\equiv t$
mod $2\pi/\Omega$ 
so that
$(X,\theta )\in I\!\! R^2 \times S^1$, and $\Sigma_{\theta_0}$
denotes 
the section
$\theta =\theta_0$.  
We consider a Hamiltonian of the form
\begin{equation}
\tilde{H}(X,t) = \tilde{H}_0 (X) + \varepsilon \tilde{H}_1 (X,t)\;\;\; 
(\tilde{H}_1 \; 2\pi\! /\Omega\! -\!\mbox{periodic in }t ),
\label{11}
\end{equation}
where $\tilde{H}_0$ is integrable with homoclinic (heteroclinic)
loop $\Gamma$ associated to
some {\it hyperbolic} fixed saddle point(s). Under hyperbolicity
hypothesis it can be 
shown that
for sufficiently small $\varepsilon$ the invariant
manifolds are only
 deformed, and possibly
their intersections become transversal, see for
instance Ref. \cite{guckenheimer}.

Let $X_0$ be a point on $\Gamma$ and
$X^{s/u} (\theta_0 ,\varepsilon )$ be 
points on the
stable/unstable manifolds such that they
are on $\Sigma_{\theta_{0}}$, in
 the line
perpendicular to $\Gamma\times \{\theta_0 \}$ at $X_0$ and whose 
trajectories take the
least amount of time to reach/depart any small neighborhood of 
the unstable periodic orbit. A computable measure
of the transversal distance between the stable and unstable 
manifolds on $\Sigma_{\theta_{0}}$,
which defines the Melnikov function, is given by the zero
 order term of
$1/\varepsilon [\tilde{H}_0 (X^u (\theta_0 ,\varepsilon )) -
\tilde{H}_0 (X^s (\theta_0 ,\varepsilon ))]$ \cite{ozorio}.
In fact, if
$X^{s/u}(t;\theta_0 ,\varepsilon )$ denotes the time 
evolution under $\tilde{H}$
such that $X^{s/u}(\theta_0;\theta_0 ,\varepsilon ) =
 X^{s/u}(\theta_0 ,\varepsilon )$,
and $X_0 (t)$ denotes the time evolution under
$\tilde{H}_0$ such that
$X_0 (0)=X_0$,
\begin{eqnarray}
\tilde{H}_0 (X^{s/u} (\theta_0 , \varepsilon ))
 -\tilde{H}_0 (X^{s/u} (\pm\infty ;\theta_0 , \varepsilon ))=
\int_{\pm\infty}^{\theta_0} \frac{d\tilde{H}_0}{dt}\left[ X^{s/u}
(t;\theta_0 , \varepsilon )\right] dt \nonumber\\
=\int_{\pm\infty}^{\theta_0}
\frac{d\tilde{H}_0}{dt}\left[ X_0 (t-\theta_0 ), t\right]
 dt + O(\varepsilon^2).
\label{12}
\end{eqnarray}
Thus in the homoclinic case the Melnikov function
can be written as
\begin{equation}
M(\theta_0)=\frac{1}{\varepsilon}\int_{-\infty}^{\infty}
 \frac{d\tilde{H}_0}{dt}\left[ X_0 (t),t+\theta_0\right] dt ,
\;\;\;\left(\frac{1}{\varepsilon}\frac{d\tilde{H}_0 }{dt}=
\{\tilde{H}_0 ,\tilde{H}_1\}\right) ,
\label{13}
\end{equation}
where $\{\;.\;,\;.\;\}$ are the usual Poisson brackets.

The implicit function theorem allows us to conclude
 that if $M(\theta_0 )$ has simple zeros,
then, for sufficiently small $\varepsilon$ the invariant
 manifolds intersect transversely for
some $\theta_0$. On the other hand, if $M(\theta_0)$ is
 bounded away from zero, then
the invariant manifolds do not intersect for all $\theta_0$.

Now let us take the map defined by system (\ref{3})-(\ref{4}) on 
a arbitrary section. After a
scale change in the $p$ coordinate it reads
\begin{eqnarray}
u_{k+1} = u_k - C u_k^3 \left[ p_k + O(4) \right],
\label{14}\\
p_{k+1} = p_k - C u_k^3 \left[ u_k + O(3) \right],
\label{15}
\end{eqnarray}
where $C =\sqrt{2}\pi/\Omega$. Here the standard Melnikov
 method breaks down because of the
degeneracy of the saddle point. McGehee \cite{gehee},
Xia \cite{xia},
 Dankowicz and Holmes
\cite{dankowicz} studied systems of this class in the context
 of the three body problem, where
they established the fundamental results needed
to support the Melnikov method:
structural stability of the unstable periodic orbits (trivial in our
case since the degenerated unstable periodic orbit
remains fixed); existence of local stable and unstable
analytic manifolds $C^{\infty}$ close to those of the
 unperturbed case \cite{gehee,dankowicz};
solutions on the perturbed and the unperturbed manifolds 
approach to the unstable periodic orbit at a similar rate
\cite{xia,dankowicz}. Following the proofs step-by-step we can
 observe that all the essential
hypothesis involved to achieve their results are also  satisfied
 by the above system. Thus
these statements apply to Eqs. (\ref{14})-(\ref{15}) allowing the 
expansion in Eq. (\ref{12}) that justifies the use of Melnikov method 
in the present problem.

Another important result whose standard form assumes
 hyperbolicity is the Smale-Birkhoff homoclinic theorem.
It was given a formulation of this theorem that 
is valid for the degenerated problem of
Sitnikov \cite{dankowicz}, which is grounded on a suitable
 approximation of the linearization
of the map in the neighborhood of the saddle point. Since
 such approximation results in the same
expressions for the case of Eqs. (\ref{14})-(\ref{15}), we conclude 
that the Smale-Birkhoff homoclinic theorem applies to the
above system. Thus transversal homoclinic intersections in
 our problem lead to the
Smale horseshoe.

Now we shall  apply Melnikov method to Eq. (\ref{1}) subjected to 
the perturbations (\ref{10}).
In the homoclinic case $(0\leq\beta <1/16)$ we find for
 the Melnikov function
\begin{eqnarray}
M(\theta_0 )&=&\int_{-\infty}^{+\infty}nr^{-n-1}
\cos [\Omega (t+\theta_0 )]\frac{dr}{dt}dt
\label{16}\\
&=& -2 n\sin (\Omega\theta_0 )K(\Omega ),\;\;\;\nonumber\\
 K(\Omega )&\equiv&\int_{0}^{v_-}v ^{n-1}\sin [\Omega t(v)]dv,
\label{17}
\end{eqnarray}
where we have fixed $X_0 = [v(0),p(0)]$ as the
symmetry point on the loop
and the integrand $t(v)$ means the positive
branch of Eqs. (\ref{5})-(\ref{6}).
Thus the Melnikov function has simple zeros as
long as $K(\Omega )\neq 0$.
With the change $t\to v$ we pass from an infinite
interval in Eq. (\ref{16})
 to a finite one
in Eq. (\ref{17}), and it allows us to study
$K(\Omega)$ using graphics.
Although $u$ ($r = 1/u^2$) is important
to justify the Melnikov method,
we have great freedom in the choice of a
coordinate to study the results.
The most simple one that is adequate to
this end is the coordinate $v=1/r$.

The integrand of $K(\Omega )$ is formed by the product of an
 oscillating function and a
polynomial. Near the origin this oscillation is a rapid 
one, since $t\longrightarrow \infty$
as the particle goes to the unstable periodic orbit. In Fig. 4 we show a
 graph of $\sin [\Omega t(v)]$
for $\beta = 1/25$, and several values of  $\Omega$. For
 $\Omega >4$ we will have more zeros in the interval
shown in the figure.
For $\Omega <1$ the curve will look like the one for
$\Omega =1$. Since
 the area under
the curves are clearly not null the integral of
$\sin [\Omega t(v)]$
is nonzero for $\beta =1/25$ and $0<\Omega\leq 4$. The cases of
 interest are $n\geq 2$, where we
will have a more favorable situation.
We  have transverse homoclinic orbits
in all these cases.
To better understand this behavior we show in Fig. 5
a graph of the integrand of $K(\Omega )$
for $\beta =1/25$, $\Omega =3$ and different values of $n$.
For $0\leq\beta <1/16$ the graphics of
$\sin [\Omega t(v)]$ will look like
the one for $\beta = 1/25$, with almost
the same upper bound for $\Omega$. See for instance
Fig. 6, where we plot the graphics of
$\sin [\Omega t(v)]$ for the monopolar attraction center
($\beta = 0$) for the same values of
$\Omega $ employed in Fig. 4. Therefore 
we have transverse homoclinic orbits
for all $\beta$ limited 
by $0\leq\beta <1/16$.

For these range of parameters the Smale-Birkhoff
 homoclinic theorem implies 
the existence of a hyperbolic
invariant set for which the action of an $N$th iterate of the
 map has a symbolic dynamic
equivalent to that of the Smale horseshoe map. Some important
 consequences of
this result are sensitive dependence on initial condition 
(a characteristic of chaos);
nonexistence of real analytic integral of
motion (nonintegrable system);
existence of infinitely many periodic
orbits with arbitrary large periods (whose number
increases exponentially with the period);
capture of orbits by the system
(in both directions of time).

It is illustrative  to see  how the chaotic orbits look
in the original noncompactified coordinates $r,p$.
 Due to the lack of an extra integral
of motion  the particle can have a highly erratic motion 
 and  have  access to a two dimensional region of the phase space.
 The sensitive
dependence on initial condition implies that the
evolution of two infinitesimally near points in the
space $r,p$ can result in two completely different
bounded orbits, in two completely different unbounded
orbits, or even in one bounded and one unbounded orbit.
Orbits that are bounded for all $t<0$ can go to infinite
for $t\longrightarrow +\infty$, and orbits of particles
coming from infinite can remain bounded for all $t>0$.
However, regular orbits are also present and
in particular there is a family of periodic orbits.  

Let us consider the heteroclinic case $\beta = 1/16$.
The Melnikov method obtained from
Eq. (\ref{12}) applies to each branch of the loop. The distance between
$X^{s/u} (\pm\infty ;\theta_0 , \varepsilon )$
and $X^{s/u} (\pm\infty ;\theta_0 , 0)$
is of order $O(\varepsilon )$. Therefore
$H_0 (X^{s/u} (\pm\infty ;\theta_0 , \varepsilon ))$
is of order $O(\varepsilon ^2)$.
Then, this term can be neglected
and the Melnikov function reads
\begin{eqnarray}
M(\theta_0 )= 2 n\sin (\Omega\theta_0 )K_s (\Omega ) - 2 n\cos
 (\Omega\theta_0 )K_c (\Omega ),
\label{18}\\
K_s (\Omega )\equiv\int_{0}^{v_-}v ^{n-1}\sin [\Omega t(v)]dv, \;\;
K_c (\Omega )\equiv\int_{0}^{v_-}v ^{n-1}\cos [\Omega t(v)]dv.
\label{19}
\end{eqnarray}
The orbits in the heteroclinic loop are less symmetrical than in the
 homoclinic one
and, consequently, the Melnikov function has simple zeros when at
 least one of the integrals of
the previous formula is different from zero. The integrands 
of $K_s$ and $K_c$ oscillate very
rapidly near the unstable periodic orbits. To better understand this
 behavior we 
show in Fig. 7 a graphic of
the positive branch of Eq. (\ref{8}). The function $t(v)$ has small 
values for a large
range of values of $v$, and hence $K_c$ will result nonzero 
values for small $\Omega $.
Indeed, we show in Fig. 8 a graph of $\cos [\Omega t(v)]$ for
different values of $\Omega$. Since 
cosine is an even function,
the same figure is valid for the negative branch.
For $\Omega >3$ we will
have more zeros in the interval $1<v<4$.
For $\Omega <0.5$ the curves 
will look like the
one for $\Omega =0.5$. For small values of
$n$ the integral $K_c$ will be nonzero
for $0<\Omega\leq 3$.
Due to the change of sign of $\cos [\Omega t(v)]$
near $v=4$, the upper bound for
$\Omega $ decrease with $n$. But it is clear
from Fig. 8 that for each $n$ will exist
a upper bound $\Omega_o (n) > 0$ such that
$K_c$ will be nonzero,
leading to the presence of transversal heteroclinic orbits,
for $0<\Omega\leq\Omega_o (n)$.

\section{Poincar\'e sections method}

The system (\ref{9}) has four parameters, $\beta$, $n$, $\Omega$,
and $\varepsilon$. In opposition to the Melnikov method, Poincar\'e 
sections method is able to predict results only for fixed  values
of  these parameters. However, the Poincar\'e method can
locate the regular and irregular regions and provide a
qualitative idea of chaotic behavior.

The perturbed phase space is in $I\!\! R^2 \times S^1$
and the maps defined on different values of the angular
variable $\theta = t$ mod $2\pi/\Omega$ are topologically
conjugated. So we have restricted the study to sections built
on $\theta =0$, but for a large number of different
values of the parameters $\beta$, $n$, $\Omega$, and $\varepsilon$.
We included $\Omega $ values which would require a more elaborate
numerical computation of the Melnikov function.
In Fig. 9 we show the Poincar\'e sections for
some select values of $\beta$, $n$, $\Omega$, and $\varepsilon$.
The general aspect of the sections is  represented in
this figure, with the predominance of irregular behavior
near the destroyed invariant loop and the
presence of a regular region near the center.

A careful analysis of the Poincar\'e sections
reveals that chaos increases with $n$, which is natural
since the major contribution of the perturbation comes
from $v>1$. In Figs. 9(a) and 9(b) we show
the Poincar\'e sections for $n=2$ and $n=5$, respectively,
where $\beta = 1/25$, $\Omega =1$, and $\varepsilon =0.003$. 
As a function of $\Omega$, the most chaotic behavior 
occurs for frequencies of the same order of the angular
frequencies of the orbits in the local minimum of $V_{eff}$,
from $(4/3)^2$ for $\beta =1/16$ to $1$ for $\beta =0$.
It is a reasonable result since for small $\Omega$ the
system is almost autonomous, and for large frequencies the 
particle feels only an average of the multipolar motion
that goes to zero for $\Omega\longrightarrow\infty$.
In Figs. 9(a), 9(c), and 9(d) we show  Poincar\'e
sections for $\Omega =1$, $\Omega = 0.1$, and $\Omega = 5$,
respectively, where $\beta = 1/25$, $n=2$, and
$\varepsilon = 0.003$.
Also, the chaos increases strongly with the ``size'' of the
perturbation $\varepsilon$. In Figs. 9(a) and 9(e)
we show the Poincar\'e sections for $\varepsilon = 0.003$
and $\varepsilon =0.01$, respectively,
where $\beta = 1/25$, $n=2$, and $\Omega = 1$.

The form of the orbits change with $\beta$, since there is a
change in the destroyed loop, but the relative
chaotic area of the section is almost independent of the
value of $\beta$. In Figs. 9(a), 9(f), and 9(g)
we show the Poincar\'e sections for $\beta =1/25$, $\beta = 0$
and $\beta = 1/16$, respectively, where $n=2$, $\Omega = 1$
and $\varepsilon =0.003$.
In particular, the case $\beta =1/16$ presents two unstable
periodic orbits and no higher chaotic behavior seems to be
associated to it.

\section{Final remarks}

Making use of a compactification procedure and extensions of the 
standard Melnikov method and Smale-Birkhoff homoclinic theorem,
we have detected transverse homoclinic
and heteroclinic orbits and  Smale horseshoe
in periodically perturbed monopole
$+$ quadrupolelike potential.
Moreover the compactified coordinates
have proved useful to work out the
Poincar\'e section method.

Although we have studied only one particular system,
we stress that the approach is general
and can be applied to a large class of open systems.
The mathematical results depend
essentially on the presence of the monopolar term
in the unperturbed equations.
Thus the same approach and with the same
compactification applies to every
inner-multipolar expansion model with
nonzero monopolar contribution.
In such cases we can allow non-Hamiltonian
perturbations of the form $f(u,t,\varepsilon )$
in the equation corresponding to Eq. (\ref{4}),
where $f$ is an  analytic function of its
variables, periodic in $t$, satisfying
$f(u,t,0)=0$ and of order $O(u^6)$.

These systems represent very general situations.
In the gravitational case, for instance,
they can model a potential due to a mass
distribution moving periodically with
reflection symmetry. This includes all
planar potential written as Fourier expansion
in time together with an inner-multipolar
expansion in space variable. A specific example is
provided by a Newtonian binary system perturbed
by gravitational radiation.
This system models the long-term dynamical
evolution of binary systems of stars
due to the emission and absorption of
gravitational radiation \cite{chicone}.

In electromagnetic systems we can have a different situation.
The possible absence of
monopolar contribution creates a difficulty in
transforming the map equations into the analytic
form (\ref{14})-(\ref{15}). The solution may be found using
 the Cassasayas-Fontich-Nunes's
result, which establishes the Melnikov method for
systems with parabolic degenerated saddle
points \cite{fontich}. A suitable
compactification to treat the electromagnetic case
is under study.

\vskip 0.5truecm

\leftline{ACKNOWLEDGMENTS}

The authors are thankful to CNPq and 
Fapesp for financial support.



\newpage

\begin{figure}[h]
\begin{center}
\psfig{file=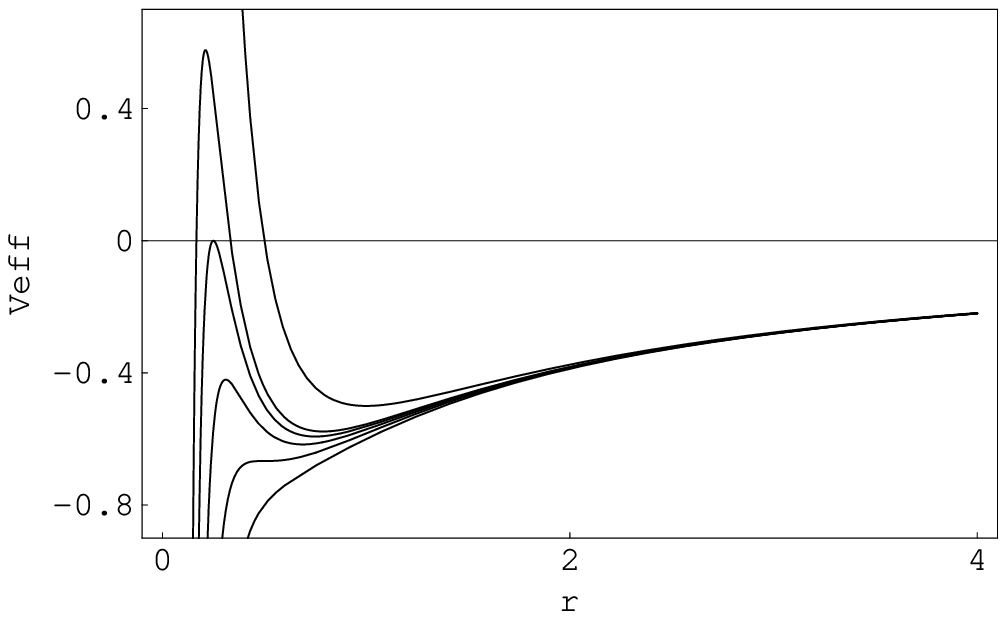}
\caption{The effective potential $[V_{eff}(r)]$ for
$\beta =1/10$ (bottom curve), $\beta =1/12$,
$\beta =1/14$, $\beta =1/16$, $\beta =1/18$, and
$\beta = 0$ (top curve), where the last one
represents a monopolar potential.}
\label{fig1}
\end{center}
\end{figure}

\begin{figure}[h]
\begin{center}
\psfig{file=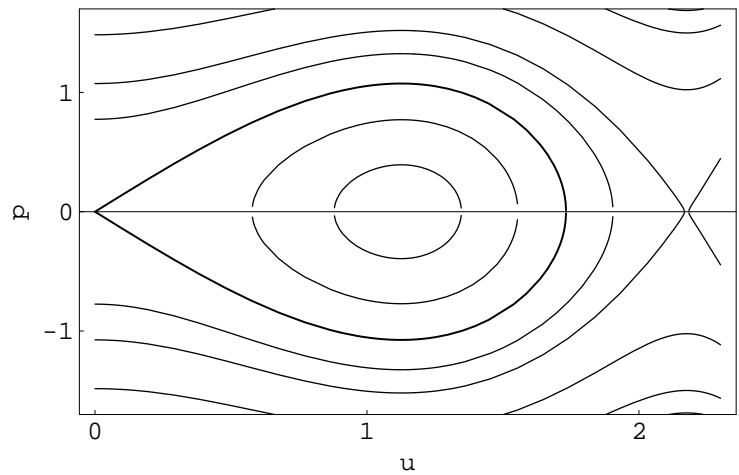}
\caption{The level curves of the Hamiltonian
$H_0 $ for $\beta =1/18$.
The  homoclinic loop associated to $(u,p) = (0,0)$  is the
 curve that contains that point.}
\label{fig2}
\end{center}
\end{figure}

\begin{figure}[h]
\begin{center}
\psfig{file=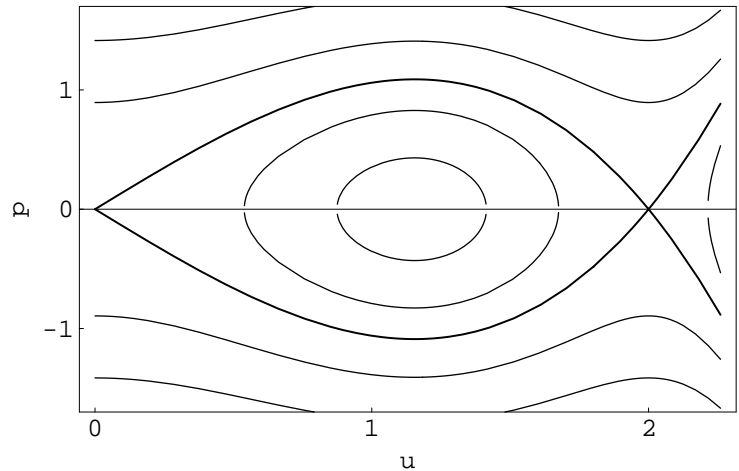}
\caption{The level curves of the Hamiltonian
$H_0 $ for $\beta =1/16$.
The  curves that represent the invariant
manifolds, which define the
heteroclinic loop, associated to the points
$(u,p) = (0,0)$ and $(u,p)= (2,0)$ are the curves
 that contains those  points.}
\label{fig3}
\end{center}
\end{figure}

\begin{figure}[h]
\begin{center}
\psfig{file=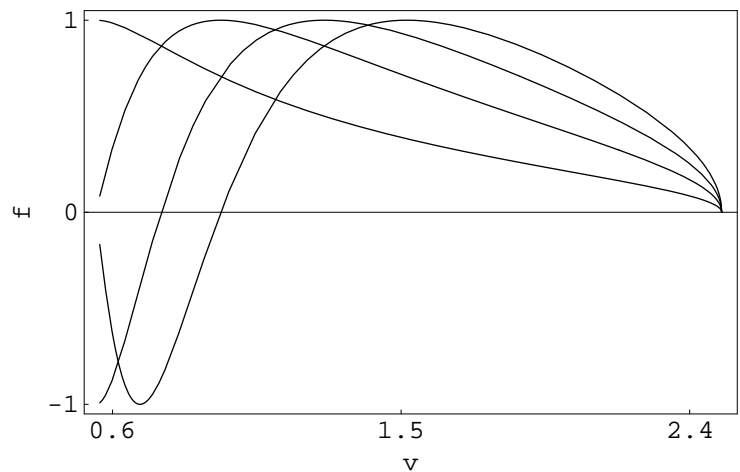}
\caption{$f\equiv\sin [\Omega t(v)]$ for
$\beta = 1/25$, and $\Omega =1$ (bottom curve),
$\Omega =2$, $\Omega =3$, and $\Omega =4$ (top curve).}
\label{fig4}
\end{center}
\end{figure}

\begin{figure}[h]
\begin{center}
\psfig{file=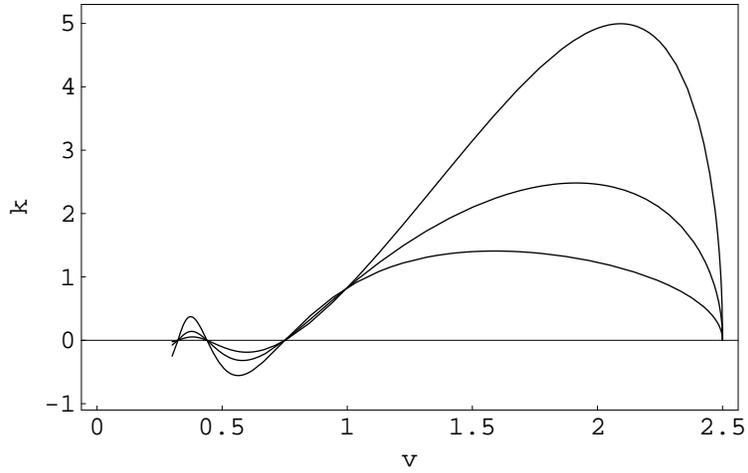}
\caption{The integrand $k\equiv v^{n-1}\sin [\Omega t(v)]$
of $K(\Omega )$ for $\beta =1/25$, $\Omega =3$, and $n=2$
(bottom curve), $n=3$ and $n=4$ (top curve).} 
\label{fig5}
\end{center}
\end{figure}

\begin{figure}[h]
\begin{center}
\psfig{file=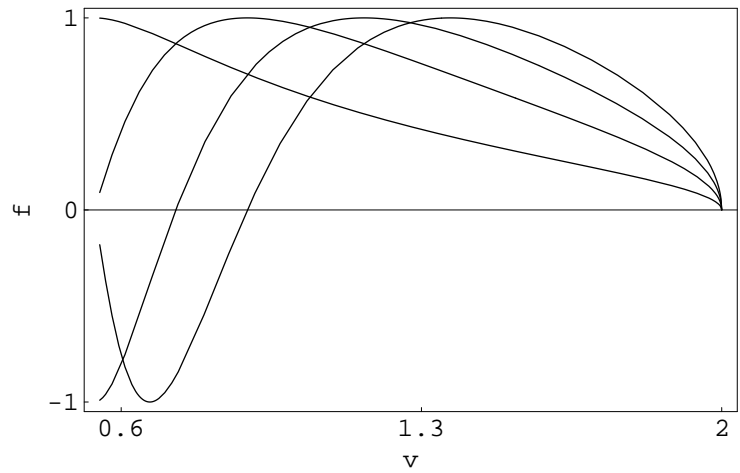}
\caption{$f\equiv\sin [\Omega t(v)]$ for
the monopolar potential
($\beta = 0$) with $\Omega =1$ (bottom curve),
$\Omega =2$, $\Omega =3$, and $\Omega =4$ (top curve).}
\label{fig6}
\end{center}
\end{figure}

\begin{figure}[h]
\begin{center}
\psfig{file=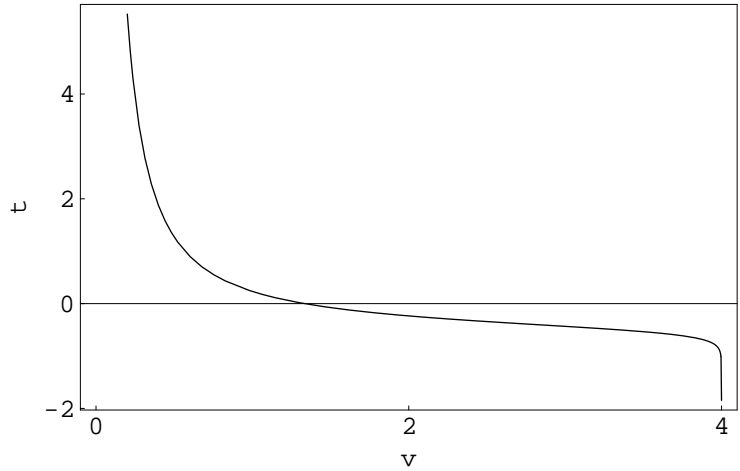}
\caption{The positive branch of Eq. (8): $t(v)$
goes to $\pm\infty$ at $v=0$ and $v=4$,
but has small values at almost every point
of the interval $0<v<4$. }
\label{fig7}
\end{center}
\end{figure}

\begin{figure}[h]
\begin{center}
\psfig{file=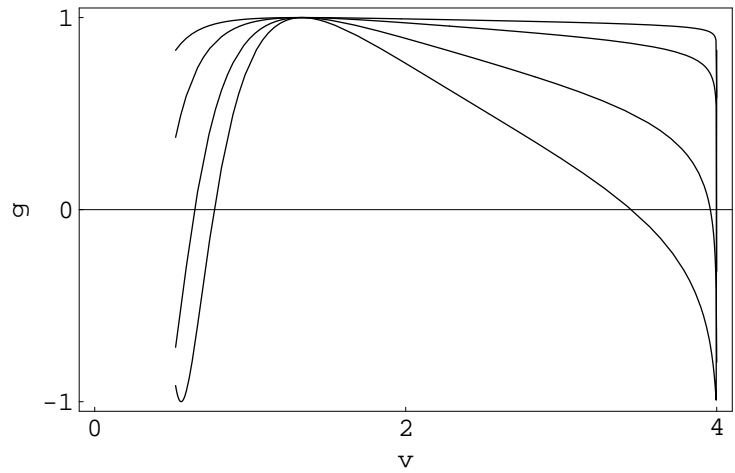}
\caption{$g\equiv\cos [\Omega t(v)]$ for
$\beta = 1/16$, and $\Omega =0.5$ (top curve),
$\Omega =1$, $\Omega =2$, and $\Omega =3$ (bottom curve).}
\label{fig8}
\end{center}
\end{figure}

\begin{figure}[h]
\begin{center}
\psfig{file=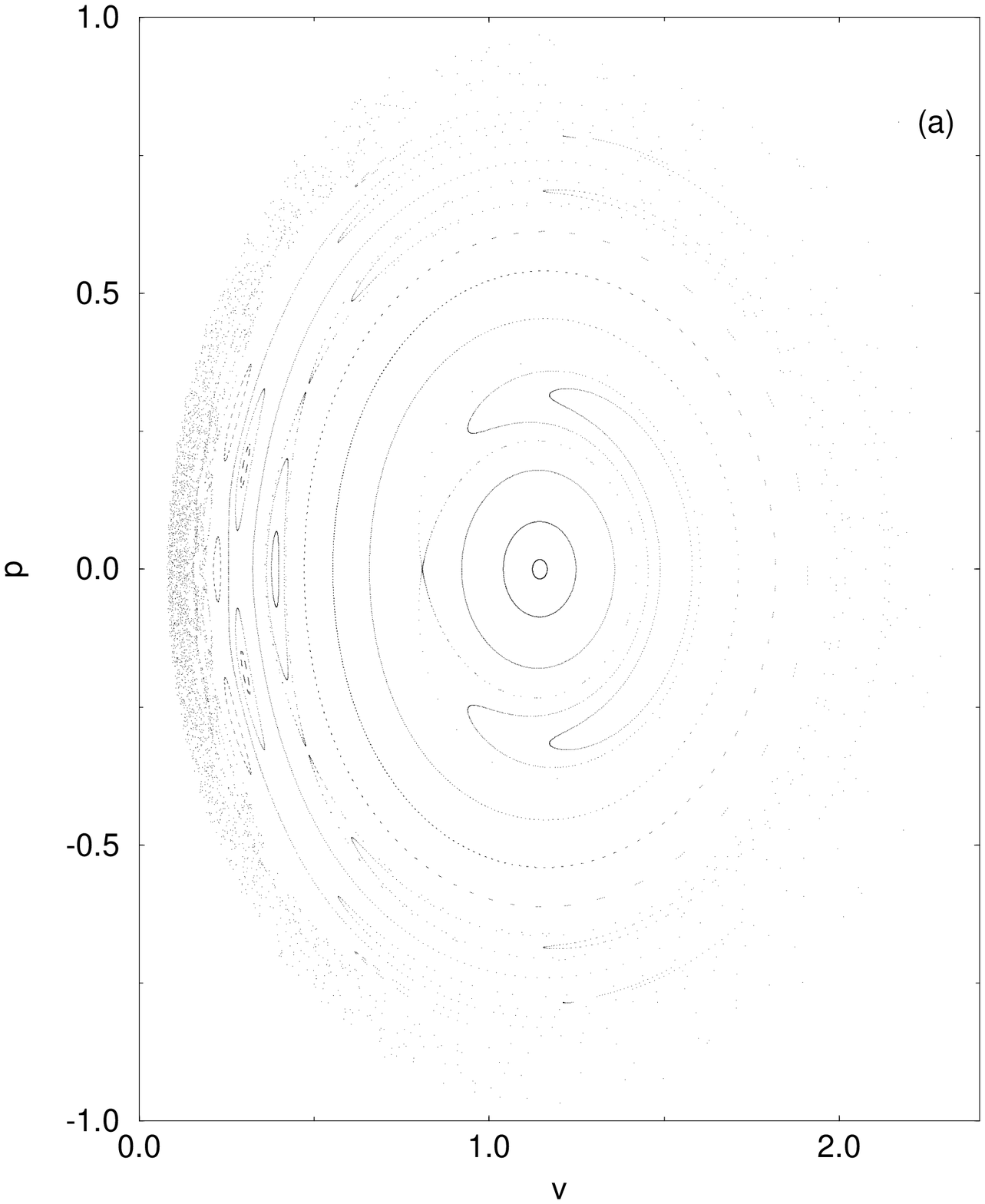, width=6cm}
\hspace{2cm}
\psfig{file=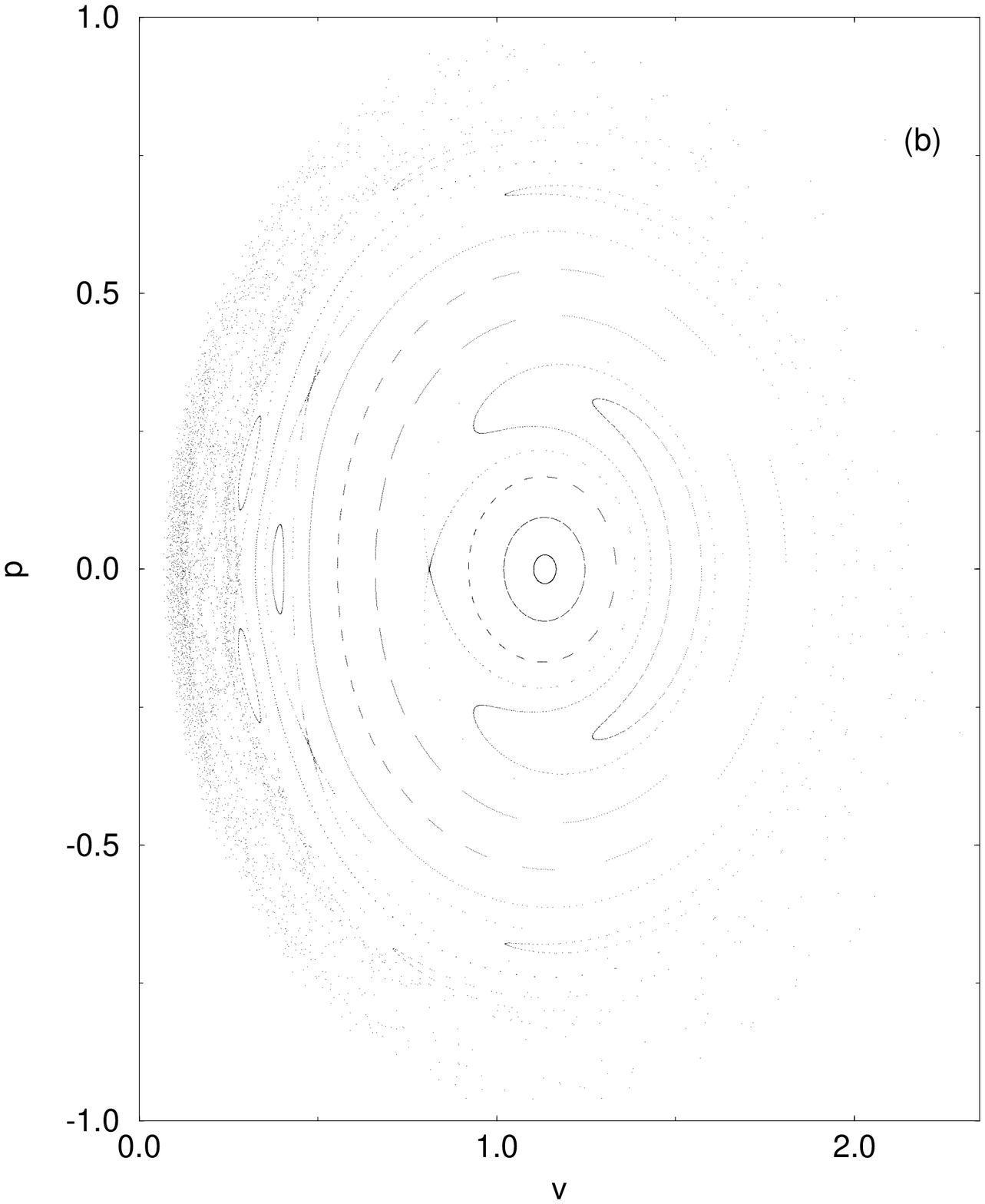, width=6cm}

\vspace{2 cm}
\psfig{file=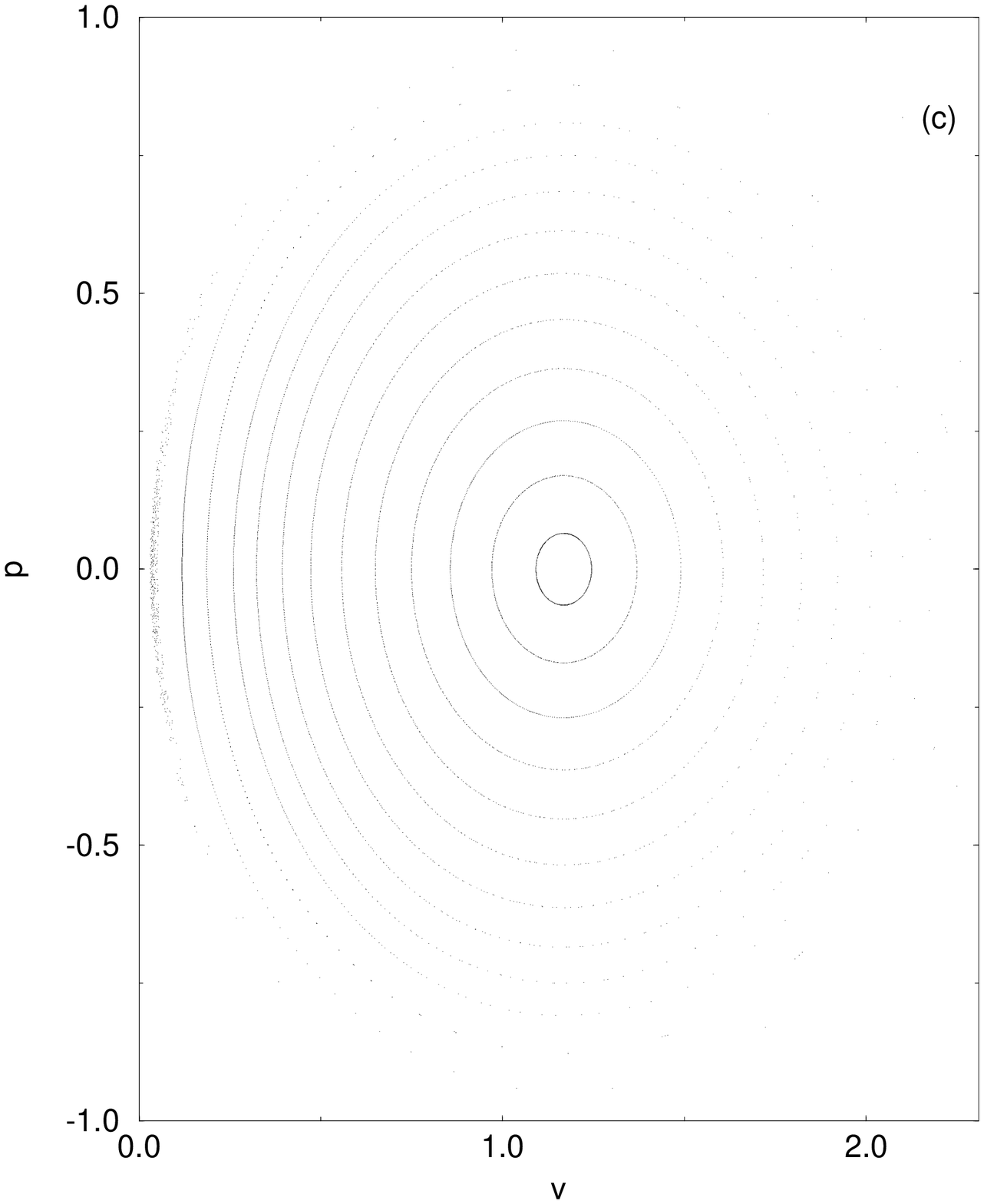, width=6cm}
\hspace{2cm}
\psfig{file=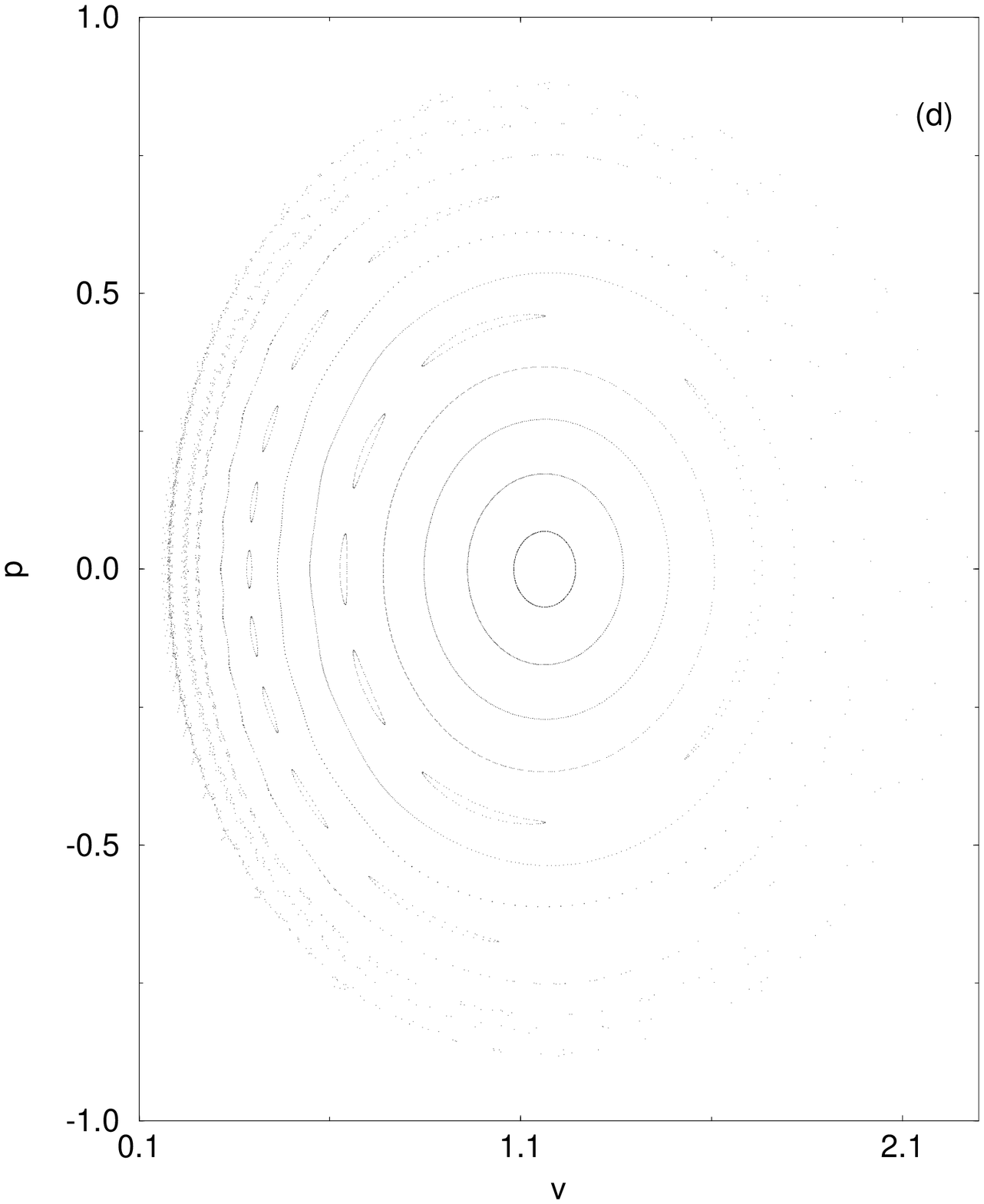, width=6cm}
\end{center}
\caption{The Poincar\'e sections for
(a) $\beta =1/25$, $n=2$, $\Omega =1$, and $\varepsilon = 0.003$;
(b) $\beta =1/25$, $n=5$, $\Omega =1$, and $\varepsilon = 0.003$;
(c) $\beta =1/25$, $n=2$, $\Omega =0.1$, and $\varepsilon = 0.003$;
(d) $\beta =1/25$, $n=2$, $\Omega =5$, and $\varepsilon = 0.003$;
(e) $\beta =1/25$, $n=2$, $\Omega =1$, and $\varepsilon = 0.01$;
(f) $\beta =0$, $n=2$, $\Omega =1$, and $\varepsilon = 0.003$;
(g) $\beta =1/16$, $n=2$, $\Omega =1$, and $\varepsilon = 0.003$.}
\label{fig9}
\end{figure}

\setcounter{figure}{8}
\begin{figure}[h]
\begin{center}
\psfig{file=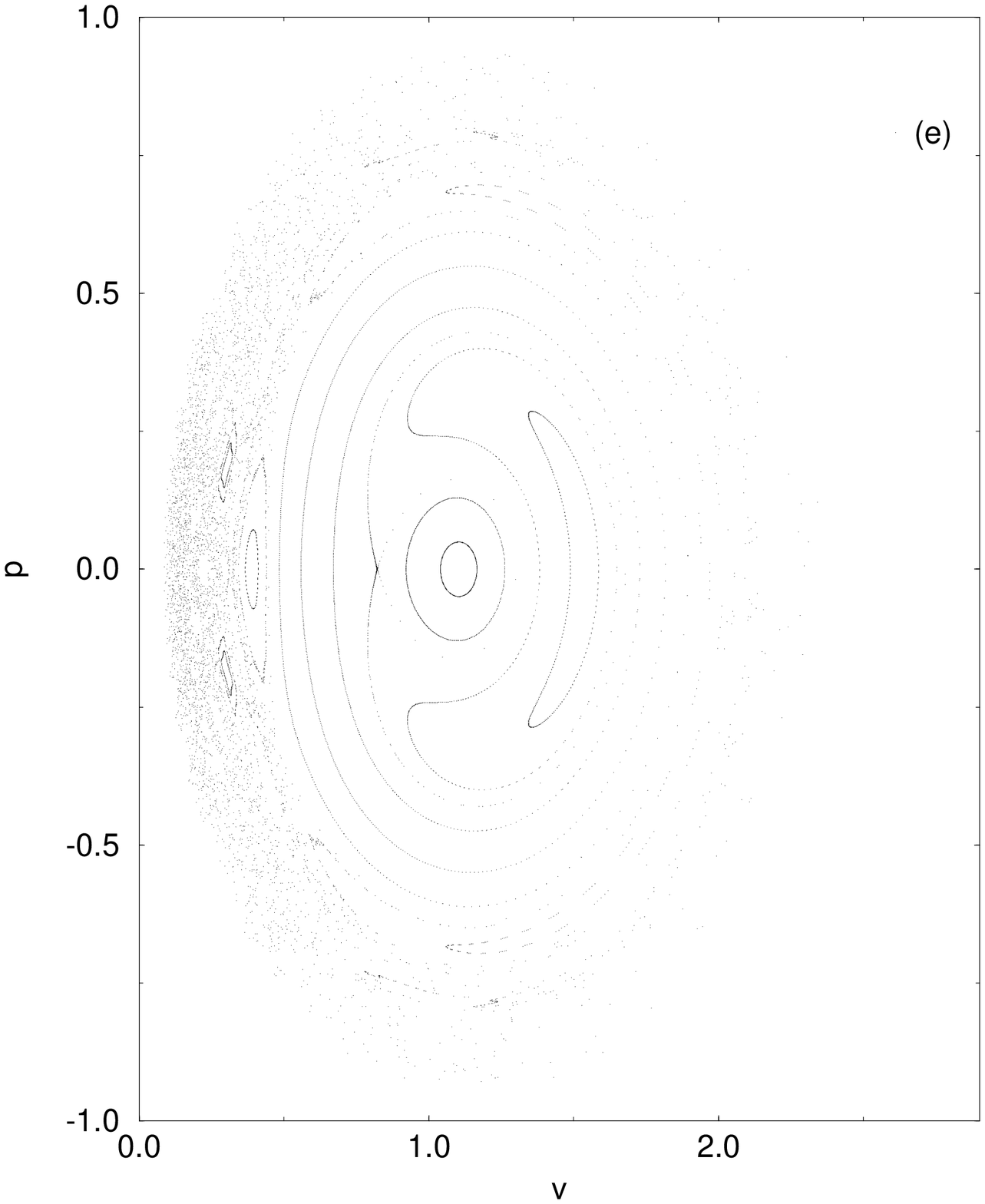, width=6cm}
\hspace{2cm}
\psfig{file=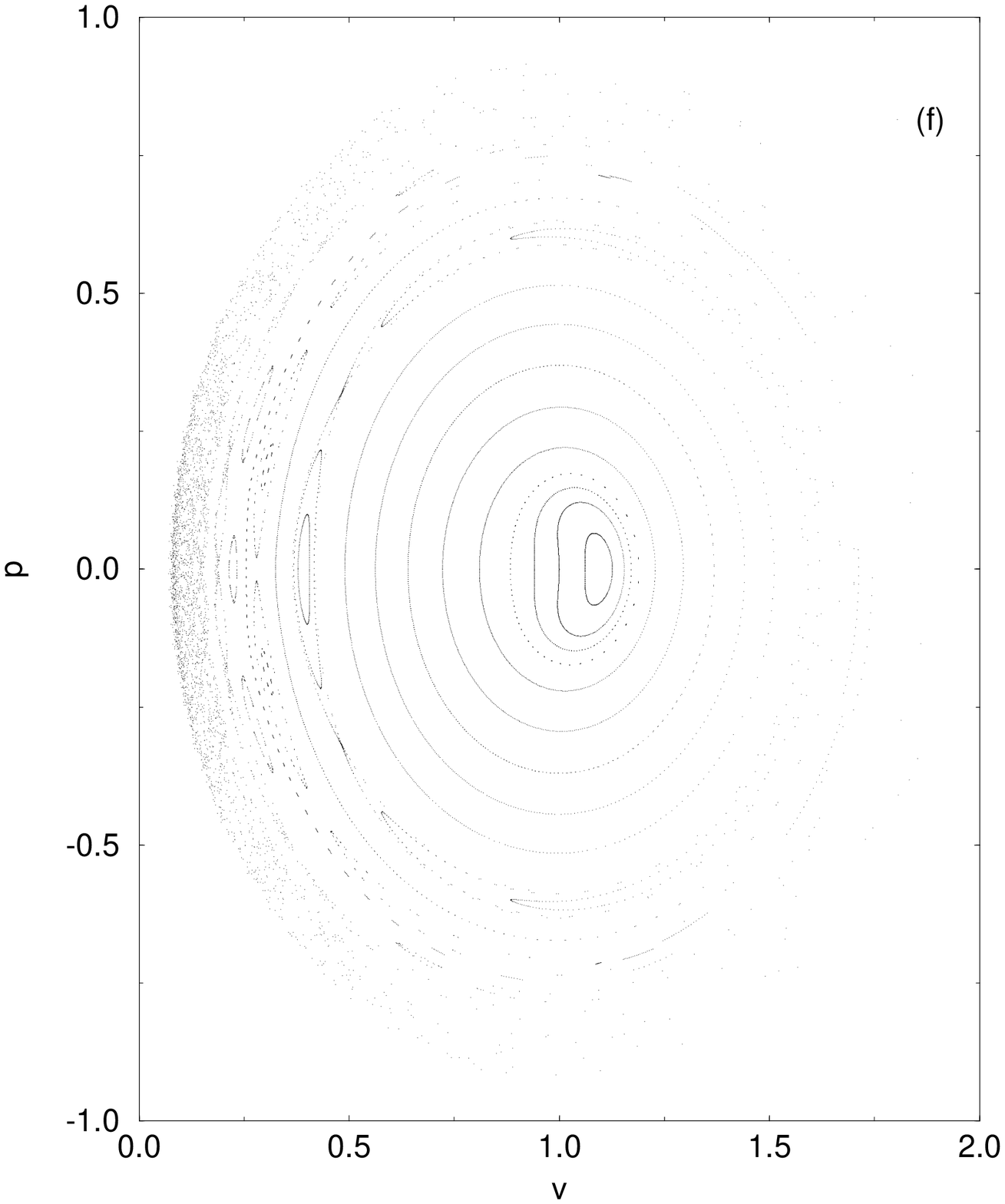, width=6cm}

\vspace{2 cm}
\psfig{file=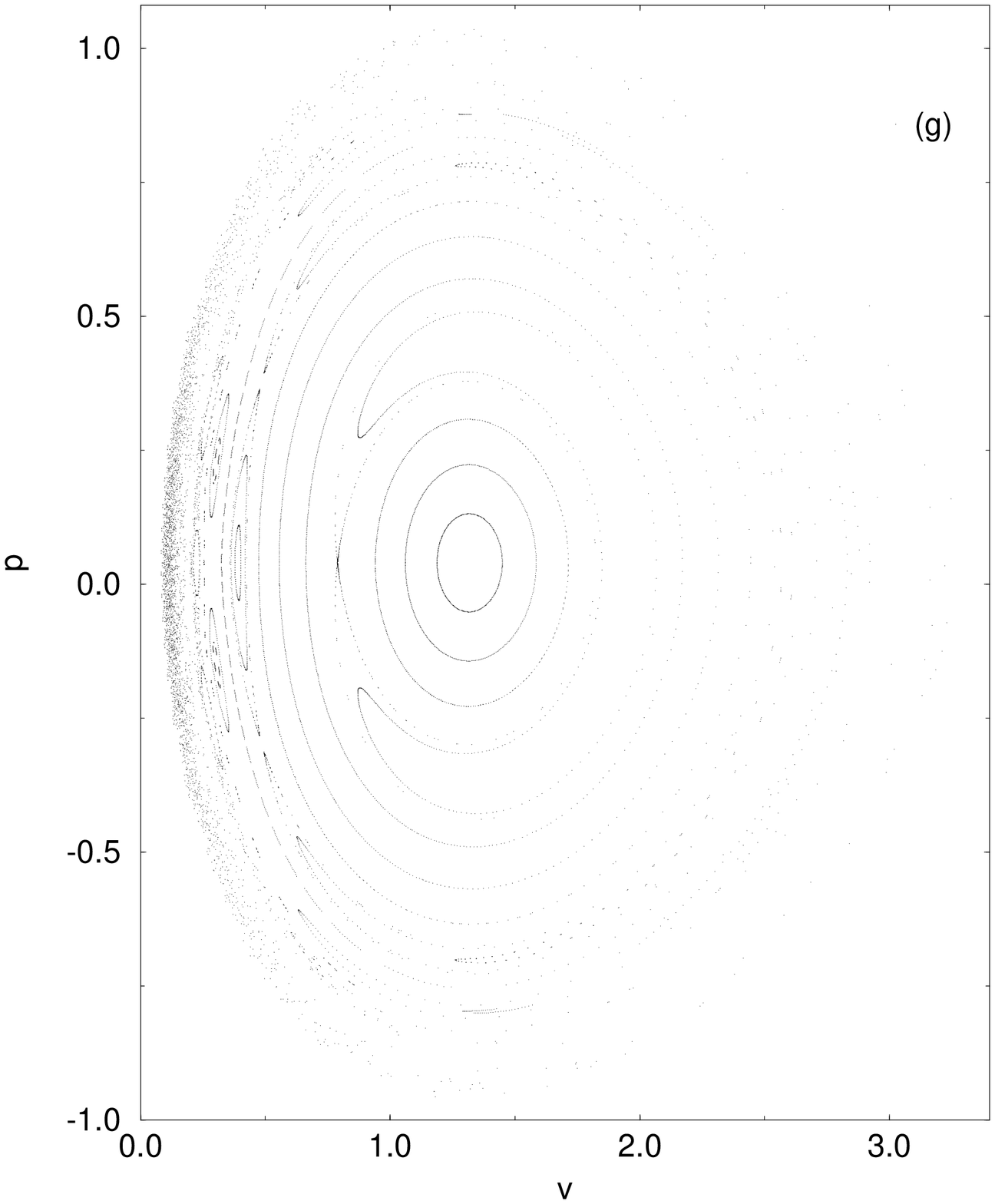, width=6cm}
\end{center}
\caption{(Continued)}
\end{figure}

\end{document}